\documentstyle[12pt]{article}
\date{}
\def\be{\begin{equation}}
\def\ee{\end{equation}}
\def\bea{\begin{eqnarray}}
\def\eea{\end{eqnarray}}
\def\s{\sigma}
\def\al{\alpha}

\def\de{\delta}
\def\om{\omega}
\def\pr{\prime}

\title{Closed relativistic string carrying pointlike mass}
\author{A.\,E. Milovidov, G.\,S. Sharov\\
{\small Tver state university}\\
{\small Tver, 170002, Sadovyj per. 35, Mathem. dep-t.}}
\begin{document}
\maketitle
\begin{abstract}
For the closed relativistic string carrying a pointlike mass
the exact solutions of the dynamical equations are obtained and studied.
These solutions describe states of the mentioned system
moving in Minkowski space and also in the space that is the direct
product of Minkowski space and a compact manifold (torus).
\end{abstract}


We consider the closed relativistic string with one massive point
(homeomorphic to a circle with a marked point) moving in the space
$M=R^{1,3}\times K$. Here $R^{1,3}$ is $3+1$\,-\,dimensional Minkowski space
and $K$ is a compact manifold with $D-4$ dimensions resulting from the
compactification procedure of the string theory \cite{GSW}.

The action for the considered system is
\be
S=-\gamma\int\limits_{\Omega}\sqrt{-g}\;d\tau d\s
-m\int\sqrt{\dot x_1^2(\tau)}\;d\tau.
\label{S}\ee

Here $\gamma$ is the string tension,
$m$ is the pointlike mass,
$g$ is the determinant of the induced metric
$g_{ab}=G_{\mu\nu}(X)\,\partial_aX^\mu\partial_bX^\nu$
on the string world surface $X^\mu(\tau,\s)$, embedded in $M$,
the speed of light $c=1$,
$\Omega=\{\tau,\s:\,\tau_1<\tau<\tau_2,\,\s_1(\tau)<\s<\s_2(\tau)\}$;
the equations $x_i^\mu(\tau)=X^\mu(\tau,\s_i(\tau))$,
$i=1,2$ describe the same trajectory of the string massive point
\be
x_1^\mu(\tau)=X^\mu(\tau,\s_1(\tau))\stackrel{\sim}{=}X^\mu(\tau^*,\s_2(\tau^*))
\label{close}\ee
on the world surface that is tube-like \cite{Tr,PRTr}.
These equations may contain two different parameters $\tau$ and $\tau^*$, connected via the relation
$\tau^*=\tau^*(\tau)$. This relation should be included in the closure condition
(\ref{close}) of the world surface.
The symbol of equality
$\stackrel{\sim}{=}$ in the condition (\ref{close})
will be defined later.

Equations of motion of this system result from the action (\ref{S}) via its
variation. They may be reduced to the simplest form
under the orthonormality conditions on the world surface
\be
(\partial_\tau X\pm\partial_\s X)^2=0,
\label{ort}\ee
and the conditions
\be
\s_1(\tau)=0,\qquad \s_2(\tau)=2\pi.
\label{ends}\ee
Conditions (\ref{ort}), (\ref{ends}) always may be fixed without
loss of generality, if we choose the relevant coordinates $\tau$, $\s$
\cite{PRTr,stab}.
The scalar square in Eq.~(\ref{ort})
results from scalar product
$(\xi,\zeta)=G_{\mu\nu}\xi^\mu\zeta^\nu$.

The orthonormality conditions (\ref{ort}) are equivalent to the conformal
flatness of the induced metric $g_{ab}$.
Under these conditions the string motion equations take the form
\be
\frac{\partial^2X^\mu}{\partial\tau^2}-
\frac{\partial^2X^\mu}{\partial\s^2}+\Gamma^\mu_{\lambda\nu}(X)
\left(\frac{\partial X^\lambda}{\partial\tau}\frac{\partial X^\nu}{\partial\tau}-
\frac{\partial X^\lambda}{\partial\s}\frac{\partial X^\nu}{\partial\s}\right)=0.
\label{eqg}\ee

In this paper we consider the compact manifold $K$ as the
$D-4$\,-\,dimensional torus: $K=S^1\times S^1\times\dots\times S^1$ \cite{GSW}.
Assume that the torus $K$ and the manifold $M=R^{1,3}\times K$ are flat
and introduce coordinates $x^\mu$ so that $x^0$, $x^1$, $x^2$, $x^3$
are in $R^{1,3}$, but other coordinates (we shall denote them $x^k$, $k=4,5\dots$ below) describe $K$. These coordinates $x^k$ are cyclical
that is points with their values $x^k$ and  $x^k+N\ell_k$, $N\in Z$
are identified: $x^k\stackrel{\sim}{=}x^k+N\ell_k$.

This factorization is used in
the equality in the closure condition (\ref{close}).
Taking into account equalities (\ref{ends}),
we'll rewrite the condition (\ref{close}) in the form
\be
X^\mu(\tau^*,2\pi)=X^\mu(\tau,0)+\sum\limits_{k}N_k\ell_k\de^\mu_k.
\label{cl}\ee
Here $\de^\mu_k=\left\{\begin{array}{cl} 1, & \mu=k\\ 0, & \mu\ne k.
\end{array}\right.$

The metric of this manifold $M$ coincides with
the metric od Minkowski space $R^{1,D-1}$: $G_{\mu\nu}=\eta_{\mu\nu}
={}\mbox{diag}\,(1;-1;\dots;-1)$. Hence, in Eq.~(\ref{eqg})
$\Gamma^\mu_{\lambda\nu}=0$ and this equation becomes linear.

But the equation of motion for the massive point $m$ resulting from the action
(\ref{S}) (it may be interpreted as the boundary condition) remains nonlinear:
\be
m\frac d{d\tau}\frac{\dot X^\mu(\tau,0)}{\sqrt{\dot X^2(\tau,0)}}+\gamma
\big[X^{'\!\mu}(\tau^*,2\pi)-X^{'\!\mu}(\tau,0)\big]=0.
\label{kr}\ee
Here
$\dot X^\mu\equiv\partial_\tau X^\mu$, $X^{'\!\mu}\equiv\partial_\s X^\mu$.

Thus, dynamics of the closed relativistic string with a material point
in the manifold $M$ is described by the system (\ref{ort})\,--\,(\ref{kr}).

We search solutions of this system using the approach supposed in
Ref.~\cite{PeSh} for the string with massive ends.
On the considered world surfaces in the frameworks of the orthonormality
gauge (\ref{ort}) one can choose the coordinates $\tau$, $\s$
resulting in linearization of boundary conditions (\ref{cl}), (\ref{kr})
under the following restrictions:
\be
\sqrt{\dot X^2(\tau,0)}=C={}\mbox{const},\qquad
\tau^*=\tau+\tau_0,\qquad\tau_0={}\mbox{const}
\label{linear}\ee
Then the equations of motion are reduced to the system
\be
\begin{array}{c}\displaystyle
\frac{\partial^2X^\mu}{\partial\tau^2}-\frac{\partial^2X^\mu}{\partial\s^2}
=0,\\
X^\mu(\tau+\tau_0,2\pi)=X^\mu(\tau,0)+\sum\limits_{k}N_k\ell_k\de^\mu_k,\rule[3mm]{0mm}{1mm}\\
\ddot X^\mu(\tau,0)+Q\big[X^{'\!\mu}(\tau+\tau_0,2\pi)-X^{'\!\mu}(\tau,0)\big]=0,
\end{array}
\label{zq}
\ee
including also the orthonormality conditions (\ref{ort}).
Here
$$Q=\frac{\gamma C}m.$$

The solution of system (\ref{zq}) is obtained with the help of Fourier method
in the form of linear combination of the terms
$X^\mu(\tau,\s)=\al^\mu T(\tau)\,u(\s)$.

System (\ref{zq}) results in two boundary-value problems for functions
$T(\tau)$ and $u(\s)$:
 \be
T''(\tau)+\lambda T=0,\qquad T(\tau+\tau_0)=\kappa T(\tau);
\label{T}\ee
\be
u''(\s)+\lambda u=0,\quad u(2\pi)=\kappa^{-1} u(0),
\quad \lambda u(0)=Q\big[\kappa u'(2\pi)-u'(0)\big].
\label{u}\ee
Here $N_k=0$ is assumed in the closure condition (\ref{cl}).

These boundary-value problems have nontrivial solutions if
eigenvalues $\lambda\ge0$.

For $\lambda=0$ the solution of the system (\ref{zq}) is the sum
of terms
$$e^\mu_k(a_k\tau+b_k\s),$$
where the closure condition (\ref{cl}) yields the relations
\be
2\pi b_k=\left\{\begin{array}{cl} -a_k\tau_0, & k\le3,\\
-a_k\tau_0+\ell_kN_k, & k>3.\end{array}\right.
\label{bk}\ee

The unit vectors
$$e^\mu_0,\;e^\mu_1,\;e^\mu_2,\,\dots\, e^\mu_{D-1}$$
form the orthonormal basis in the manifold $M$.

In the case $\lambda\equiv\om^2>0$ the parameter $\kappa$ in
Eqs.~(\ref{T}), (\ref{u}) takes the value
$\kappa=e^{\pm i\om\tau_0}$, and boundary-value problem (\ref{u})
has nontrivial solutions only under the following condition:
\be
\cos2\pi\om-\cos\om\tau_0 = \frac{\om}{2Q}\sin2\pi\om.
\label{con}\ee
These solutions
$u_n(\s)=\sin\om_n(2\pi-\s)+e^{\mp i\om_n\tau_0}\sin\om_n\s$
are eigenfunctions of the problem (\ref{u}) corresponding to the
roots $\om=\om_n$ of equation (\ref{con}).

Using the obtained eigenfunctions of problems (\ref{T}) and (\ref{u})
we represent the solution of the system (\ref{zq}) in the form of
Fourier series
\be
\begin{array}{c}\displaystyle
X^\mu(\tau,\s){}=e_0^\mu a_0\Big(\tau-\frac{\tau_0}{2\pi}\s\Big)
+\sum\limits_{k\ge3}e^\mu_k(a_k\tau+b_k\s)+\\
\displaystyle
+\sum\limits_{n=-\infty}^{\infty}\al_n^\mu
\big[\sin\om_n(2\pi-\s)+e^{i\om_n\tau_0}\sin\om_n\s\big]\,e^{-i\om_n\tau}.
\end{array}
\label{solve}\ee
Here $\om_{-n}=-\om_n$, $\al^{*\mu}_n=-\al^\mu_{-n}$ is assumed.

Series (\ref{solve}) has to satisfy the orthonormality conditions
(\ref{ort}) and equality (\ref{linear}) $\dot X^2(\tau,0)=C^2$.
This results in the following system of equations for
coefficients of this series:
\be
\begin{array}{c}\displaystyle
a_0^2\Big(1+\frac{\tau_0^2}{4\pi^2}\Big)=\sum\limits_{k\ge3}(a_k^2+b_k^2)
-4\sum_{n>0}\om_n^2(\al_n,\al_n^*)(1-\cos\om_n\tau_0\cdot\cos2\pi\om_n),\\
\displaystyle
\frac{\tau_0}{2\pi}a_0^2+\sum\limits_{k\ge3}a_kb_k+2\sum\limits_{n>0}\om_n^2
(\al_n,\al_n^*)\,\sin\om_n\tau_0\cdot\sin2\pi\om_n=0,\rule{0mm}{6mm}\\
\displaystyle
a_0^2-\sum\limits\limits_{k\ge3}a_k^2+2\sum\limits_{n>0}\om_n^2(\al_n,\al_n^*)
\sin^22\pi\om_n=C^2,\rule{0mm}{5mm}\\
 N_k(e_k,\al_n)=(e_0,\al_n)=0,\quad(\al_m,\al_n)=0,\;\;m\ne-n.
\rule{0mm}{4mm}
\end{array}
\label{syst}\ee

System of equations (\ref{con}), (\ref{syst}) determine values of
parameters $\tau_0$ and $\om_n$ for the series (\ref{solve}).

We consider below the solutions (\ref{solve}) containing only one nonzero
frequency $|\om_n|$ (one-frequency solutions).
The simplest of them are generalization of well known uniform rotations
(rotational motions) of the rectilinear string in Minkowski space $R^{1,3}$
\cite{PeSh,4B,InSh}.
In the manifold $M$ (unlike in $R^{1,3}$) these solutions are nontrivial
even in the simplest case $\tau_0=0$ when equation (\ref{con}) takes the form
\be
\sin\pi\om\cdot\Big(\sin\pi\om+\frac{\om}{2Q}\cos\pi\om\Big)=0.
\label{q}\ee
The roots $\om_n=n$ of equation (\ref{q}) yield nontrivial solutions
\be
X^\mu=e_0^\mu a_0\tau
+\sum_{k>3}e^\mu_k b_k\s+A_n\sin n\s\cdot e^\mu(n\tau),
\label{sl}\ee
where $e^\mu(n\tau)=e^\mu_1\cos n\tau+e^\mu_2\sin n\tau$,
$a_0=\sqrt{n^2A_n^2+\sum b_k^2}$, $b_k=\ell_kN_k/(2\pi)$. It is assumed
$a_k=0$ for $k\ge3$ without loss of generality because the terms with
$a_k$ in Eq.~(\ref{solve}) describe a uniform rectilinear motion of the system
at a constant velocity. It may be eliminated via the Lorentz transformation.

Solutions (\ref{sl}) describe the string uniformly rotating in the plane
$e_1,\,e_2$. This closed string has the form of sinusoid with a finite length
because of cyclical nature of coordinates $x_k$, $k>3$. The massive point
is at rest at $\s=0$ (or we have the case $m=0$).

Other roots of Eq.~(\ref{q}) (roots of the equation
$\tan\pi\om=-\frac{\om}{2Q}$)
are dimensionless frequencies of the following rotational motions:
\be
X^\mu=e_0^\mu a_0\tau
+\sum_{k\ge3}e^\mu_k b_k\s+A_n\cos\big[\om_n(\s-\pi)\big]
\cdot e^\mu(\om_n\tau).
\label{rot}\ee
They are similar to motions (\ref{sl}), but the mass $m$ (at the point $\s=0$)
in the case (\ref{rot}) moves along the circle with rotating string.
Two string segments (joined at a non-flat angle) hold this mass.

In the case $a_k=b_k=0$ ($\forall k\ge3$)
expressions (\ref{sl}) and (\ref{rot}) describe
uniformly rotating closed string folded in two (then it may be folded
in $n-1$ times) in Minkowski space $R^{1,3}$.
Such string states were classified in Ref.~\cite{PeSh}
by the example of the open string with massive ends. The string in these states
has points moving at the speed of light, and the world surface has peculiarities
of the metric $\dot X^2=X^{'2}=0$ at these points' world lines.

But when the cyclical degrees of freedom are excited ($b_k\ne0$)
solutions (\ref{sl}) and (\ref{rot}) have no such peculiarities,
all string points move slower than the speed of light.

Let us consider the one-frequency solutions (\ref{solve}) in more general case
with nonzero values of the parameter $\tau_0$. In the proper system of reference
(where $a_k=0$) these solutions are represented in the form
\be
\begin{array}{c}\displaystyle
X^\mu=e_0^\mu a_0\Big(\tau-\frac{\tau_0}{2\pi}\s\Big)
+\sum\limits_{k>3}e^\mu_k\frac{\ell_kN_k}{2\pi}\s+\\
+A_n\Big\{\big[\sin\om_n(2\pi-\s)
+\cos\om_n\tau_0\cdot\sin\om_n\s\big]\cdot
e^\mu(\om_n\tau)-\sin\om_n\tau_0\cdot\sin\om_n\s\cdot\acute e^\mu(\om_n\tau)
\Big\}.
\rule[3mm]{0mm}{1mm}
\end{array}
\label{hyp}\ee
Here $\acute e^\mu(\om_n\tau)=-e^\mu_1\sin\om_n\tau+e^\mu_2\cos\om_n\tau$.
Values $\tau_0$, $\om_n$, $a_0$, $A_n=\sqrt{-2(\al_n,\al_n^*)}$
are determined from the system (\ref{con}), (\ref{syst});
$\al_n^\mu=\frac12A_n(e^\mu_1+ie^\mu_2)$. The coefficients $b_k$ are
defined by Eq.~ (\ref{bk}).

If $a_k=b_k=N_k=0$ ($\forall k\ge3$), the first two equations (\ref{syst})
result in the relation
\be
\frac{\tau_0}{2\pi}+\frac{2\pi}{\tau_0}=
2\frac{1-\cos\om_n\tau_0\cdot\cos2\pi\om_n}
{\sin\om_n\tau_0\cdot\sin2\pi\om_n}.\\
\label{omtau}\ee

In this case system (\ref{con}), (\ref{omtau}) determines the spectrum of values
$\tau_0$ and $\om_n$ without dependence on amplitude factors $a_0$ and $A_n$.

In the case $a_k=b_k=0$ solutions (\ref{hyp}) describe
uniform rotations of the string that has the form of
the closed hypocycloid joined at non-zero angle in the massive point
(this form is the section $t={}$const of the world surface (\ref{hyp})).
Hypocycloid is the curve drawing by a point of a circle (with radius
$r$) rolling inside another fixed circle with larger radius
$R$ \cite{Tr,PRTr}. In the case (\ref{hyp}) the relation of these radii
$$
\frac rR=\frac{1-|\theta|}2
$$
is irrational if $m\ne0$ and $\tau_0\ne0$. Here and below
$$
\theta=\frac{\tau_0}{2\pi}.
$$

This hypocycloid rotates in the $e_1,\,e_2$
plane at the angular velocity $\Omega=\omega_n/a_0$
where $a_0$ is determined from equations (\ref{syst}):
\be
a_0^2=\frac{\sin\om_n\tau_0}{\sin\om_n\tau_0-\theta\sin2\pi\om_n}
\bigg(\frac{mQ}\gamma\bigg)^2=\frac{A_n^2\om_n^2}{\theta}
\sin\om_n\tau_0\cdot\sin2\pi\om_n.
\label{a0}\ee
The massive point moves at the speed $v=A_n\sin2\pi\om_n$
along the circle with radius $v/\Omega$.
There are also cusps (return points)
of the hypocycloid moving at the speed of light.
There are many topologically different types of hypocycloidal solutions
(\ref{hyp}). They may be classified with the number of cusps and the
type of intersections of the hypocycloid. Note that in the considered
model (\ref{S}) the string does not interact with itself in a point of
intersection.

The solutions similar to (\ref{hyp}) with $b_k=0$ were obtained first in
Ref.~\cite{Tr} for the string baryon model ``triangle".

The more general solutions (\ref{hyp}) including the summands with $b_k\ne0$
(for cyclical coordinates in $M$) differ from mentioned hypocycloidal solutions:
their sections $t={}$const are spatial (not flat) curves, closed because of
cyclical nature of coordinates $x_k$. Their projections onto the plane
$e_1,\,e_2$ look like hypocycloids. But these world surfaces have no
peculiarities of the metric $\dot X^2=X^{'2}=0$.

\smallskip

Possible applications of the obtained solution for the considered model
include describing of intrinsic degrees of freedom for an elementary
particle with the given set of quantum numbers and parameters.
Let us calculate the most important among them:
the energy $E$ and angular momentum $J$ for the one-frequency solutions
(\ref{solve}) of this model.
For an arbitrary classic state of the relativistic string
with the action (\ref{S}) they are determined by the following integrals
(Noether currents) \cite{PRTr,BN,osc}:
\bea
&\displaystyle
P^\mu=\int\limits_c p^\mu(\tau,\s)\,d\s
+p_1^\mu(\tau),\qquad
p^\mu(\tau,\s)=\gamma\frac{(\dot X,X') X^{\pr\mu}-X'{}^2\dot X^\mu}
{\big[(\dot X,X')^2-\dot X^2X^{\pr2}\big]^{1/2}},& \label{Pimp}\\
&\displaystyle
{\cal J}^{\mu\nu}=\int\limits_c\Big[X^\mu(\tau,\s)\,
p^\nu(\tau,\s)-X^\nu(\tau,\s)\,p^\mu(\tau,\s)\Big]\,d\s+
\Big[x_1^\mu(\tau)\,p_1^\nu(\tau)-x_1^\nu(\tau)\,p_1^\mu(\tau)\Big],&
\label{Mom}\eea
where $x_1^\mu(\tau)=X^\mu\big(\tau,\s_1(\tau)\big)=X^\mu(\tau,0)$ and
$p_1^\mu(\tau)=m\dot x_1^\mu(\tau)\big/\sqrt{\dot x_1^2(\tau)}$ are
coordinates and momentum of the massive point, $c$ is any closed curve
(contour) on the tube-like world surface of the string.
Note that the lines $\tau={}$const on the world surface (\ref{hyp})
are not closed in the case $\tau_0\ne0$. So we can use the most suitable
lines $\tau-\theta\s={}$const (that is $t={}$const) as the contour $c$
in integrals (\ref{Pimp}), (\ref{Mom}).

The reparametrization $\tilde\tau=\tau-\theta\s$, $\tilde\s=\s -\theta\tau$
keeps the orthonormality conditions (\ref{ort}). Under them
$p^\mu(\tau,\s)=\gamma\dot X^\mu(\tau,\s)$.

The square of energy $E^2$ equals the scalar square of the conserved
vector of momentum (\ref{Pimp}): $P^2=P_\mu P^\mu=E^2$.
If we substitute the expression (\ref{hyp}) (generalized for the case $a_k\ne0$)
into Eq.~(\ref{Pimp})
we'll obtain the following formula for the momentum:
\be
P^\mu=\gamma\bigg\{a_0\Big[2\pi(1-\theta^2)+Q^{-1}\Big]\,e_0^\mu+
\sum\limits_{k\ge3}\Big[2\pi a_k(1-\theta^2)+a_kQ^{-1}+\theta\ell_kN_k\Big]
\,e_k^\mu \bigg\}.
\label{P}\ee

In the simplest case $a_k=0$, $N_k=0$ this expression takes the form
\be
P^\mu=\bigg[2\pi\gamma a_0(1-\theta^2)+\frac m{\sqrt{1-v^2}}\bigg]\,e_0^\mu.
\label{P0}\ee
Here $v$ is the constant speed of the massive point.

The classical angular momentum (\ref{Mom}) is not conserved value
for the considered system because of
anisotropy of the space $M$. Only the components ${\cal J}^{\mu\nu}$
with $\mu,\nu=0,1,2,3$ (relating to the space $R^{1,3}$) are conserved.
Among them only $z$-component of the angular momentum is nonzero:
\be\begin{array}{c}
{\cal J}^{\mu\nu}= j_3^{\mu\nu}J,\rule[-3mm]{0mm}{1mm}\\
\displaystyle
J=\frac{\gamma}{2\om_n}\left\{2\pi\bigg[a_0^2(1-\theta^2)-
\sum_{k\ge3}\bigg(a_k^2(1-\theta^2)+\frac{\ell_k^2N_k^2}{4\pi^2}\bigg)\bigg]
+\frac1Q\bigg(a_0^2-\sum\limits_{k\ge3}a_k^2-\frac{m^2Q^2}{\gamma^2}\bigg)
\right\}. \end{array}
\label{J}\ee
Here $j_3^{\mu\nu}=e_1^\mu e_2^\nu-e_1^\nu e_2^\mu=e^\mu\acute e^\nu-
e^\nu\acute e^\mu.$

In the proper system of reference with $a_k=0$ for given cyclical numbers
$N_k$ the states (\ref{hyp}) are determined by the parameters
$a_0$, $\om_n$, $\tau_0$ (or $\theta=\frac{\tau_0}{2\pi}$).
If the values $m$, $\gamma$ and the topological type of the hypocycloidal
solution (\ref{hyp}) are fixed we obtain the one-parameter set of motions
with different values $E$ and $J$. These states lay at quasilinear Regge
trajectories.

In particular, in the case $N_k=0$ one can use $Q$ as the mentioned
parameter of this set and for every value $Q$ determine $\om_n$ and
$\tau_0$ from the system of transcendental equations (\ref{con}), (\ref{omtau}) and then the amplitude factors $a_0$, $A_n$ from Eqs.~(\ref{a0}).
The values $E$ (\ref{P0}) and $J$ (\ref{J}) in this case are connected
by the relation
$$J=\frac1{2\Omega}\Big(E-m\sqrt{1-v^2}\Big)
=\frac{a_0}{2\om_n}\Big(E-m\sqrt{1-v^2}\Big).
$$

The mentioned Regge trajectories are nonlinear for small $E$ and tend to linear
if $E\to\infty$. Their slope in this limit depends on the values $N_k$ and
the fixed topological type.

\end{document}